\documentclass[12pt,draft,superscriptaddress,nofootinbib]{revtex4}
\usepackage{graphicx}
\usepackage{dcolumn}
\usepackage{amsmath,amsthm,amssymb}
\usepackage{subfigure}

\def\ra{\rightarrow}

\begin{document}

\title{A remark about wave equations on the extreme Reissner-Nordstr\"om black hole
exterior}

\author{Piotr Bizo\'n}
\affiliation{Institute of Physics, Jagiellonian
University, Krak\'ow, Poland}
\affiliation{Max Planck Institute for Gravitational Physics (Albert Einstein Institute),
Golm, Germany}
\author{Helmut Friedrich}
\affiliation{Max Planck Institute for Gravitational Physics (Albert Einstein Institute),
Golm, Germany}
\date{\today}
\begin{abstract} We consider a massless scalar field propagating on the exterior of the extreme Reissner-Nordstr\"om black hole. Using a discrete conformal symmetry of this spacetime, we draw a one-to-one relationship between the behavior of the field near the future horizon and near future null infinity. In particular, we show that the polynomial  growth of the second and higher transversal derivatives along the horizon, recently found by Aretakis, reflects well known facts about the retarded time asymptotics at null infinity. We also observe that the analogous relationship holds true for an axially symmetric massless scalar field propagating on the extreme Kerr-Newman background.

\end{abstract}

\maketitle

\noindent
Recently, Aretakis  studied a massless scalar field
on the exterior of the extreme Reissner-Nordstr\"om  black hole and proved that second and higher transversal derivatives of the field grow polynomially along the horizon, provided that a certain conserved quantity on the horizon is nonzero \cite{a1,a2}. This fact was interpreted as  indicating an instability of the extreme Reissner-Nordstr\"om  black hole.
The aim of this note is to point out that an important observation by Aretakis on the behaviour of scalar fields on  the horizon  is a reflection of well known results about the asymptotic behaviour of scalar fields near null infinity.
Moreover, we show that if Aretakis'  conserved quantity vanishes, then the second transversal derivative at the horizon  is bounded but, generically,  the third and higher ones grow.

 The exterior (or domain of outer communication)
 of the extreme Reissner-Nordstr\"om black hole is the globally hyperbolic space-time with manifold
  $\mathcal{M} =  \{ -\infty < t < \infty,  \,\, 0 < r < \infty\} \times S^2$ and metric
\begin{equation}\label{rn}
g = -A^{-2}\,dt^2 + A^2\,(dr^2 + r^2\,d\omega^2), \quad \quad A = 1 + \frac{m}{r},
\end{equation}
where $m$ is a positive constant and $d\omega^2$ is the round metric on the unit two-sphere.
The metric $g$ is the unique spherically symmetric solution of the Einstein-Maxwell equations
with mass $m$ and charge $q = \pm m$. The Maxwell field is given by $F = q/(r+m)^2\, dt\wedge dr$.
Note that we are using the isotropic radial coordinate $r$ which is related to the areal radial coordinate $R$ by $r=R-m$.

Key to our discussion is the fact that the metric $g$ admits a discrete conformal symmetry~\cite{ct}, namely, the spatial  inversion $\iota: (t, r) \rightarrow (t, m^2/r)$ of $\mathcal{M}$ onto itself  (suppressing
the angular coordinates, they are unaffected by our considerations), which satisfies
\[
\iota_* g = \Omega^2\,g \quad \mbox{with} \quad \Omega= \frac{m}{r}.
\]

To see the action of this symmetry on ${\cal M}$, it will be convenient to introduce the
retarded and advanced  time coordinates $u = t - r_*$, $v = t + r_*$ with
$r_* = r + 2\,m\,\log (r/m) - m^2/r$. They satisfy $du = dt - A^2\,dr$,  $dv = dt + A^2\,dr$ and
$u \circ \iota = v$. In the coordinates $(v, r)$ the metric takes the form
\[
g = g_v  \equiv - A^{-2}\,dv^2 + 2\,dv\,dr + (r\,A)^2\, d\omega^2
= - \frac{r^2}{(m + r)^2}\,dv^2 + 2\,dv\,dr + (m + r)^2\, d\omega^2,
\]
which shows that the metric extends as a real analytic metric $g_v$ (in fact as a solution to the Einstein-Maxwell equations)
onto the extension
${\cal M}_{H^+} =   \{ -\infty < v < \infty,  \,\, - m < r < \infty\} \times S^2$ of ${\cal M}$.
The null hypersurface
${\cal H}^+ =   \{ -\infty < v < \infty,  \,\, r = 0\} \times S^2$ represents the future event  horizon for $({\cal M}, g)$.
In the coordinates $(u, r)$ the metric takes the form
\[
g = g_u  \equiv - A^{-2}\,du^2 - 2\,du\,dr + (r\,A)^2\, d\omega^2.
\]
Expressed in source coordinates $(u, r)$ and target coordinates $(v, r)$ the inversion takes the form $\iota: (u , r) \rightarrow (u, m^2/r)$ and  the relation above reads
\begin{equation}
\label{gv-pullback}
\iota_* g_v = \Omega^2\,g_u.
\end{equation}
In the coordinates $(u, \rho \equiv m^2/r)$ the metric $\hat{g}_u \equiv   \Omega^2\,g_u$
and the conformal factor $\Omega$  take the form
\[
\hat{g}_u = - A(\rho)^{-2}\,du^2 + 2\,du\,d\rho + (\rho\,A(\rho))^2\, d\omega^2,
\quad \quad
\Omega = \rho/m,
\]
which shows that $\hat{g}_u$ and $\Omega$  extend as real analytic fields
(solution to the conformal Einstein-Maxwell equations)
onto the extension
${\cal M}_{{\cal J}+} =  \{ -\infty < u < \infty,  \,\, - m <  \rho < \infty\} \times S^2$ of  ${\cal M}$.
Because $\hat{g}_u$ and $g_v$ are related on ${\cal M}$ by a diffeomorphism, the metric $\hat{g}_u$ has vanishing Ricci scalar as well.
The null hypersurface
${\cal J}^+ =   \{ -\infty < u < \infty,  \,\, \rho = 0\} \times S^2$ on which $\Omega = 0$, $d \Omega \neq 0$
represents future null infinity for $({\cal M}, g)$.

In terms of source coordinates $(u, \rho )$ and target coordinates $(v, r)$ the inversion takes on ${\cal M}$ the form $\iota: (u , \rho) \rightarrow (v = u, r = \rho)$. It follows  that $\iota$ extends to
a real analytic isometry $\iota':  ({\cal M}_{{\cal J}+}, \hat{g}_u) \rightarrow ({\cal M}_{H^+}, g_v)$
which maps ${\cal J}^+$ onto ${\cal H}^+$ and ${\cal M}$ onto itself.

\vspace{.5cm}

As a consequence, `conformally well behaved' fields on the extreme Reissner-Nordstr\"om background can hardly distinguish between ${\cal J}^+$ and ${\cal H}^+$. This is quite clear for Maxwell or Yang-Mills fields. They are governed by equations which, in four dimensions, only depend on the conformal structure.
We discuss here the slightly more subtle case of scalar fields satisfying the massless wave equation.

On a four dimensional space-time $({\cal N}, h)$ with {\it vanishing Ricci-scalar}  (the situation considered here)
the wave operator $\Box_{h} \equiv \nabla_{\mu}\,\nabla^{\mu}$ is identical with the  {\it conformally covariant wave  operator}
$L_h = \Box_{h} - 1/6\,R_h$, which  satisfies for any conformal factor $\vartheta > 0$ and any scalar field $f$
\[
L_{\vartheta^2\,h}(\vartheta^{-1}\,f) = \vartheta^{-3}\,L_h(f).
\]
If $\phi: ({\cal N}', h') \rightarrow ({\cal N}, h)$ is a space-time diffeomorphism with inverse $\psi$, it holds
\[
L_{h'}(f \circ \phi) = L_{\psi_* h'}(f) \circ \phi.
\]
Applying this in the situation described by
(\ref{gv-pullback}) gives for $f \in C^2({\cal M})$
\begin{equation}
\label{mv-mu}
L_{g_u}( \tilde{f}) = \Omega^3 \cdot L_{g_v}(f) \circ \iota
\quad \mbox{with} \quad \tilde{f} = \Omega \cdot f  \circ \iota.
\end{equation}
Thus, if $f(v, r)$ solves the wave equation near ${\cal H}^+$, where
$r \rightarrow 0$,  the function $\tilde{f}(u, r) = \frac{m}{r}\,f(u, \frac{m^2}{r})$ satisfies
that equation near ${\cal J}^+$, where $r \rightarrow \infty$. It follows that any general property of solutions to the wave equation on ${\cal M}$ near null infinity  translates into a corresponding
general property of solutions to the wave equation on ${\cal M}$ near the horizon and vice versa. A Taylor series expansion of $f(v, r)$ in terms of $r$ near ${\cal H}^+$  immediately translates into a Bondi-type expansion of $\frac{m}{r}\,f(u, \frac{m^2}{r})$ near null infinity.

This relationship  is more direct in the case of  the isometry $\iota'$ where
\begin{equation}
\label{m-rescaled-m}
L_{\hat{g}_u}\,(f \circ \iota') = L_{g_v}(f) \circ \iota' \quad \mbox{for} \quad
f \in C^2({\cal M}_{H^+}).
\end{equation}
This formula  says that  if $f(v, r)$ solves the wave equation for the `physical'  metric $g_v$ near the horizon, where $r = 0$, then $f(u, \rho)$ solves the wave equation for the conformally rescaled and extended metric  near ${\cal J}^+$ where $\rho = 0$.  The wave operator $\Box_{\hat{g}_u}$ is identical with the operator $\Box_{g_v}$ after replacing the coordinates $u$ and $\rho$ by $v$ and $r$.
Any $v$-independent quantity on ${\cal H}^+$ derived from a solution to $\Box_{g_v}f = 0$ near ${\cal H}^+$
corresponds to a $u$-independent quantity on ${\cal J}^+$ derived from a solution to
 $\Box_{\hat{g}_u}f = 0$ near ${\cal J}^+$. Any statement about the decay of solutions
 toward the future on null infinity translates into a  statement about the decay on the horizon.

These observations apply in a non-trivial way to the work of Aretakis \cite{a1, a2} and  Dain and Dotti \cite{dd}. We remark first on the smoothness of the solution.
If the solution $f$ develops in time from Cauchy data on the Cauchy hypersurface
${\cal S} = \{t = 0\}$ of $({\cal M}, g)$, its smoothness at ${\cal H}^+$ and ${\cal J}^+$ depends very much on how the data are prescribed  on ${\cal S}$
near $i^0$, where $r \rightarrow \infty$, and near $i^*$, where $r \rightarrow 0$.
To avoid subtleties, Dain and Dotti prescribe data of compact support on ${\cal S}$. This  ensures that the solution
$f$ extends smoothly to ${\cal H}^+$ and ${\cal J}^+$. Aretakis prescribes instead data on a space-like hypersurface ${\cal S}'$ which approaches $i^0$ at one end
and intersects ${\cal H}^+$ as a smooth space-like hypersurface  at the other end. The data are required to be smooth up to ${\cal H}^+ \cap {\cal S}'$ so that  the solution $f$ extends smoothly to
${\cal H}^+$.

An important step in Aretakis' work \cite{a2} is the observation that with any solution to the wave equation can be associated an infinite sequence of quantities which are conserved along the null generators of ${\cal H}^+$. It is the purpose of this note to point out that this confirms a result which has been known for a long time.

In fact, $\iota'^{-1}({\cal S}')$ being a hypersurface which behaves like a hyperboloidal hypersurface near ${\cal J}^+$, the pull-back $f \circ \iota'$ is smooth on ${\cal J}^+$ and the analysis of the early studies of the asymptotic  structure  of field at null infinity applies.
After Newman and Penrose discovered the existence of non-trivial conservation laws
on ${\cal J}^+$  for asymptotically flat solutions to Einstein's field equations  (\cite{Exton-NP-consts}, \cite{NP-consts-1}, \cite{NP-consts-2}) various attempts  were made to understand the origin and the significance of these quantities
(\cite{Goldberg-N-P-const:1967}, \cite{Goldberg-N-P-const:1968}, \cite{Robinson-NP-consts}). In this context it was observed for the first time  that an infinite number of conserved quantities can be defined at ${\cal J}^+$ for linear massless fields
on asymptotically flat backgrounds (provided the solutions are sufficiently smooth on ${\cal J}^+$).

\vspace{.3cm}

Let us  illustrate the above general considerations and some of their consequences by the simple example of spherically symmetric evolution.
In this case the wave equation $L_{g_u} \tilde{f} =0$ reads (for convenience we set $m=1$)
\begin{equation}\label{equ}
 -2 \partial_{ru}  \tilde{f} - \frac{2}{1+r} \, \partial_u  \tilde{f}
 +\frac{1}{(1+r)^2}\, \partial_r(r^2 \partial_r  \tilde{f})=0\,.
\end{equation}
For solutions that are smooth at  ${\cal J}^+$ we write the Bondi-type expansion in powers of
 $1/r$
\begin{equation}\label{bondi}
  \tilde{f}(u,r)=\frac{c_0(u)}{r}+\frac{c_1(u)}{r^2}+\frac{c_2(u)}{r^3}+\dots.
\end{equation}
Plugging this expansion into \eqref{equ} and collecting terms with the same power  of $1/r$ one gets an infinite hierarchy of linear ordinary differential equations with constant coefficients of the form
\begin{equation}\label{hierarchy}
\dot c_n=\sum_{i=0}^{n-1} (\alpha_i \dot c_i+\beta_i c_i)\,,
\end{equation}
which can be integrated one by one if one knows the `radiation field' $c_0(u)$. The first three equations read:
\begin{align}
 & \dot c_0+\dot c_1=0, \label{c1}\\
 & 2\dot c_2 +\dot c_1 - \dot c_0 +c_1=0, \label{c2}\\
 & 3\dot c_3 + \dot c_2 - \dot c_1 +\dot c_0 -2 c_1 + 3 c_2=0 \label{c3} \,.
\end{align}
Equation \eqref{c1} is the conservation law for the Newman-Penrose constant $P=c_0+c_1$ (note that in the standard expansion in inverse powers of the areal radial coordinate $R=r+1$, the Newman-Penrose constant is the coefficient of the $1/R^2$ term \cite{Goldberg-N-P-const:1968}). A precise relationship between the
 asymptotic decay of  $c_0(u)$ for $u\ra \infty$ and the fall-off of initial data near $i^0$ is a difficult problem on its own which fortunately need not concern us here  because it does not affect our argument. Let us distinguish the two cases $P\neq 0$ and $P=0$. In the first case it suffices to know that $c_0\ra 0$ as $u\ra \infty$ (in fact, it is known that $c_0(u)$ tends to zero as $1/u$ or faster \cite{gsw}). Then
  $c_1(u)\ra P$ (because $c_0+c_1=P$) and the successive integration of Eqs.\eqref{c2} and \eqref{c3} gives\footnote{By $f(u)\sim g(u)$ we mean asymptotic equivalence for $u\rightarrow \infty$.}
\begin{equation}\label{P=0}
  c_1(u)\sim P \overset{\eqref{c2}}{\implies} c_2(u)\sim -\frac{1}{2} P u \overset{\eqref{c3}}{\implies} c_3(u) \sim \frac{1}{4} P u^2\,.
\end{equation}

  In the case of $P=0$, to begin with, we need to know more about the decay of $c_0(u)$. For the purpose of the argument, let us  assume the worst possible scenario, that is the fastest possible decay one can have for generic solutions, namely $c_0(u)\sim a/u^2$, where $a$ is a constant depending on the initial data \cite{gsw, dr,bcr}. Then $c_1(u)\sim -a/u^2$ (because by assumption $c_0(u)+c_1(u)=0$) and the successive integration of Eqs.\eqref{c2} and \eqref{c3} gives
   \begin{equation}\label{Pneq0}
  c_1(u)\sim -\frac{a}{u^2} \overset{\eqref{c2}}{\implies} c_2(u)\sim C-\frac{a}{2u} \overset{\eqref{c3}}{\implies} c_3(u)\sim -C u +\frac{a}{2} \ln{u}\,,
\end{equation}
 where $C$ is an integration constant depending on the initial data\footnote{For non-trivial non-generic solutions with $c_0(u)=o(u^{-2})$  the coefficient $c_3(u)$ may be bounded but it is clear that the increase of $c_n(u)$ for some $n>3$ is inevitable.}.

   In both cases, $P=0$ and $P\neq 0$, continued integration of the system \eqref{hierarchy} for $n>3$  yields  faster and faster growing coefficients $c_n(u)$.
We wish to stress that this behaviour of the coefficients $c_n(u)$ is an inherent property of the Bondi-type expansion in any asymptotically flat spacetime, in particular the Minkowski spacetime. It basically follows from the fact that along a null outgoing cone ($u=const$) the field decays as $1/v=1/(u+2r)$ which for large~$r$ generates the geometric series of powers of $u/2r$.  This reflects the fact that the decay of solutions towards $i^+$ is \emph{not uniform} because it is \emph{slower} along ${\cal J}^+$ than along timelike directions.

It follows from the above discussion of conformal symmetry of the problem that if $ \tilde{f}(u,r)$ is a solution of Eq.\eqref{equ}, then $f(v,r)=r^{-1}  \tilde{f}(u,1/r)$ is a solution of the wave equation  $L_{g_v}  f=0$, which reads
\begin{equation}\label{eqv}
  2\partial_{rv}  f + \frac{2}{1+r}\, \partial_v  f + \frac{1}{(1+r)^2}\, \partial_r(r^2 \partial_r  f)=0\,.
\end{equation}
Correspondingly, the Bondi-type expansion \eqref{bondi} near ${\cal J}^+$ translates into the Taylor series expansion near ${\cal H}^+$
\begin{equation}\label{taylor}
  f(v,r)=c_0(v)+c_1(v)\,r+ c_2(v)\,r^2+...
\end{equation}
Thus, in spherical symmetry the conserved quantity  found by Aretakis $H_0=\partial_r  f(v,0) +  f(v,0)$ is nothing else but the Newman-Penrose constant $P$. In the case of nonzero $H_0$, his argument that $\partial_{rr}  f(v,0) \sim -H_0 v$ is exactly the same as the one given in \eqref{P=0} for $c_2(u)$. In the case of $H_0=0$ (not considered by Aretakis), we see from \eqref{Pneq0} that the second derivative is bounded (this was also observed in \cite{dd} for compactly supported data) but the third one generically diverges.
It is straightforward to generalize the above analysis to all $\ell>0$ multipoles in the spherical harmonic decomposition of the scalar field and show that the Newman-Penrose constants $P_{\ell}$ and Aretakis' constants $H_{\ell}$ are the same entities.

Let us remark that the tails of linear fields propagating on the extreme Reissner-Nordstr\"om background  were first studied in the physics literature by Bi\v c\'ak \cite{b}, who noticed that the  scattering potential has the same fall-off near the horizon and near the spatial infinity. More recently, the problem was revisited by Blaksley and Burko \cite{bb}, who gave a heuristic argument and numerical evidence that the decay rates of the tails along ${\cal J}^+$ and ${\cal H}^+$ are the same (in stark contrast to non-extreme black holes for which the tails along the horizon decay as fast as in the timelike directions \cite{gpp,dr}).

\vspace{.3cm}

We conclude this note by pointing out some consequences and a natural question related to the existence of the conformal symmetry.
Obviously, relations between the behaviour of solutions near the horizon and the behaviour of solutions  near null infinity as those outlined above for linear massless
 fields can also be established for other equations which admit a   suitable transformation law under conformal rescalings. Thus, all known results on solutions of these equations near/on  null infinity translate to statements on solutions near/on the horizon.
In particular, the analysis presented above can be generalized to nonlinear conformally invariant fields propagating on the extreme Reissner-Nordstr\"om background, for instance a Yang-Mills field \cite{bkm}. For such fields in general there do not exist any conserved quantities on ${\cal H}^+$  or ${\cal J}^+$ \cite{fs}, nonetheless  one can infer the polynomial (or logarithmic) growth in time of higher transversal derivatives by integrating a nonlinear dynamical system analogous to \eqref{hierarchy}.
Finally, in an   analysis which  requires a deeper understanding of the structure of the horizon,  it may prove useful to observe  that the background itself admits conserved quantities  on the horizon which correspond to the Newman-Penrose conserved quantities at null infinity.

There arises the natural question as to whether a relationship between ${\cal J}^+$ and ${\cal H}^+$ as described  here
is only an idiosyncrasy of the extreme Reissner-Nordstr\"om solution in four dimensions or a more general property of extreme black hole solutions.
As has already been pointed out by Couch and Torrence \cite{ct}, for the extreme Kerr-Newman family satisfying $m^2=a^2+q^2$ with nonzero specific angular momentum $a$ there does not exist a conformal symmetry of the same simplicity as $\iota$.
We found, however, that an axially symmetric massless scalar field propagating on the extreme Kerr-Newman background \emph{does} have such a symmetry. To show this, let us express  the extreme Kerr-Newman metric $g_{KN}$ in terms of `isotropic'  Boyer-Lindquist coordinates $(t,r,\vartheta,\varphi)$ (in which the radial coordinate $r$ is related to the usual radial coordinate $R$ by $r=R-m$ so that the horizon is located at $r=0$)
\begin{equation}\label{kn}
  g_{KN}=-\frac{\rho^2 r^2}{B}\,dt^2 + \frac{B \sin^2{\vartheta}}{\rho^2}\,(d\varphi-\omega dt)^2+\frac{\rho^2}{r^2}\,dr^2+\rho^2\,d\vartheta^2\,,
\end{equation}
where
$$
\rho^2=(r+m)^2+a^2 \cos^2{\vartheta},\quad B=[(r+m)^2+a^2]^2-a^2 r^2 \sin^2{\vartheta},\quad \omega=\frac{m^2+a^2+2mr}{B}\,a\,.
$$
For $a=0$ the metric \eqref{kn} reduces to \eqref{rn}. The massless wave equation $\Box_{g_{KN}} f=0$ for
 an axially symmetric field $f(t,r,\vartheta)$ takes the form
\begin{equation}\label{eq}
-  \left[\frac{((r+m)^2+a^2)^2}{r^2}-a^2 \sin^2{\vartheta}\right] \, \partial_{tt} f
+  \partial_r(r^2\, \partial_r f) + \frac{1}{\sin{\vartheta}}\, \partial_{\vartheta}(\sin{\vartheta} \, \partial_{\vartheta} f) = 0.
\end{equation}
A direct calculation shows  that this equation is invariant under the spatial inversion $r\rightarrow \dfrac{m^2+a^2}{r}$, that is if $f(t,r,\vartheta)$ is a solution, so is $\dfrac{1}{r} f\left(t,\dfrac{m^2+a^2}{r},\vartheta\right)$.
Thus, the one-to-one relationship between the behaviour of the scalar field near ${\cal J}^+$ and near ${\cal H}^+$ described above for the extreme Reissner-Nordstr\"om background, can be repeated verbatim for the entire extreme Kerr-Newman family in the case of  axially symmetric fields. As a consequence, the growth  of higher transversal derivatives along ${\cal H}^+$ follows from the known decay properties of axisymmetric fields on ${\cal J}^+$, confirming the analysis of Aretakis \cite{a3}.

\vspace{.3cm}

\vskip 0.1cm \noindent \emph{Acknowledgments:}
We thank Lars Andersson for discussions and Sergio Dain for a correspondence.
The work of PB was supported in part by the NCN Grant NN202 030740.

\end{document}